\documentclass[12pt]{iopart}
\usepackage{harvard}
\usepackage{iopams} 
\usepackage{url}
\usepackage{graphicx}
\usepackage{caption}
\usepackage{lipsum}

\usepackage{hyperref}

\begin{document}

\bibliographystyle{jphysicsB}



\vspace*{-2.5cm}
\title[Nuclear Physics Uncertainties and Chemical Evolution]{The Impact of Nuclear Physics Uncertainties on Galactic Chemical Evolution Predictions}

\author{Benoit C\^ot\'e$^{1,10,11}$, Pavel Denissenkov$^{2,10,11}$, Falk Herwig$^{2,10,11}$, Chris L. Fryer$^{3,10,11}$, Krzysztof Belczynski$^{4}$, Nicole Vassh$^{5}$, Matthew R. Mumpower$^{3,6,10}$, Jonas Lippuner$^{3,10}$,\\Marco Pignatari$^{7,10,11}$, Ashley J. Ruiter$^{8,9}$}

\vspace*{0.5cm}

\address{\hspace*{-2.5cm}$^1$Konkoly Observatory, Research Centre for Astronomy and Earth Sciences, Konkoly Thege Miklos ut\\\hspace*{-2.cm}15-17, H-1121 Budapest, Hungary}
\address{\hspace*{-2.5cm}$^2$Department of Physics and Astronomy, University of Victoria, Victoria, BC, V8W 2Y2, Canada}
\address{\hspace*{-2.5cm}$^{3}$Center for Theoretical Astrophysics, LANL, Los Alamos, NM 87545, USA}
\address{\hspace*{-2.5cm}$^{4}$Nicolaus Copernicus Astronomical Center, Polish Academy of Sciences, ul. Bartycka 18, 00-716\\\hspace*{-2.cm}Warsaw, Poland}
\address{\hspace*{-2.5cm}$^{5}$Department of Physics, University of Notre Dame, Notre Dame, Indiana 46556, USA}
\address{\hspace*{-2.5cm}$^{6}$Theoretical Division, Los Alamos National Lab, Los Alamos, NM 87545, USA}
\address{\hspace*{-2.5cm}$^{7}$E.A. Milne Centre for Astrophysics, Department of Physics \& Mathematics, University of Hull,\\\hspace*{-2.cm}Hu6 7RX, UK}
\address{\hspace*{-2.5cm}$^{8}$Research School of Astronomy and Astrophysics, Australian National University, Canberra,\\\hspace*{-2.cm}ACT 0200, Australia}
\address{\hspace*{-2.5cm}$^{9}$School of Science, University of New South Wales, Australian Defence Force Academy, Canberra,\\\hspace*{-2.cm}ACT 2600, Australia}
\address{\hspace*{-2.5cm}$^{10}$Joint Institute for Nuclear Astrophysics - Center for the Evolution of the Elements, USA}
\address{\hspace*{-2.5cm}$^{11}$NuGrid Collaboration, \url{https://nugrid.github.io}}
\ead{benoit.cote@csfk.mta.hu}
\vspace{10pt}

\begin{abstract}

Modeling the evolution of the elements in the Milky Way is a multidisciplinary and challenging task. In addition to simulating the $\sim$\,13\, billion years evolution of our Galaxy, chemical evolution simulations must keep track of the elements synthesized and ejected from every astrophysical site of interest (e.g., supernova, compact binary merger). The elemental abundances of such ejecta, which are a fundamental input for chemical evolution codes, are usually taken from theoretical nucleosynthesis calculations performed by the nuclear astrophysics community. Therefore, almost all chemical evolution predictions rely on the nuclear physics behind those calculations. In this proceedings, we highlight the impact of nuclear physics uncertainties on galactic chemical evolution predictions. We demonstrate that nuclear physics and galactic evolution uncertainties both have a significant impact on interpreting the origin of neutron-capture elements in our Solar System. Those results serve as a motivation to create and maintain collaborations between the fields of nuclear astrophysics and galaxy evolution.

\end{abstract}

%
%
%
\maketitle
%
%

\section{Introduction}

Modeling the evolution of the elements within galaxies is a challenging task because of the wide variety of physical phenomena involved. Star formation \cite{mckee07}, which triggers the chemical evolution \cite{tinsley80,gibson03,prantzos08,pagel09,matteucci14}, is regulated by stellar feedback, gas flows, and galaxy mergers \cite{baugh06,hopkins12,somerville15,naab17}. Regardless of the level of complexity used to simulate (or approximate) these processes, chemical evolution simulations rely on theoretical nucleosynthetic yields \cite{nomoto13,cristallo15,karakas16,limongi18,ritter18,battino19}. Those provide the mass of elements produced and ejected by different types of stars and astronomical events that progressively enrich galaxies throughout their evolution.

One of the challenges of simulating the chemical evolution of galaxies is to quantify the reliability of numerical predictions and to provide confidence levels in the interpretation of the origin of the elements. Indeed, because of the multiple processes occurring at different scales that need to be included in chemical evolution simulations, numerical predictions carry many sources of uncertainties \cite{romano10,cote16,simonetti19}. In particular, nucleosynthetic yields, which are typically used as simple input in chemical evolution codes, are affected by stellar evolution and nuclear physics uncertainties \cite{travaglio14,deboer17,nishimura17,fryer18,jones19}. To best interpret chemical evolution predictions, it is necessary to quantify the propagation of uncertainties, from nuclear to galactic scales.

The goal of this proceedings is to highlight the significant impact of nuclear physics uncertainties on galactic chemical evolution predictions, and on our interpretation of the origin of the elements in the Milky Way. As two illustrative examples, we focus on the contribution of rapidly accretion white dwarfs on the evolution of neutron-capture elements from Kr to Mo, via the intermediate neutron-capture ($i$) process \cite{cowan77,dardelet15,hampel16,roederer16,banerjee18,clarkson18}, and on the contribution of neutron star mergers on the evolution of Eu, via the rapid neutron-capture ($r$) process \cite{arnould07,thielemann17,cowan19,horowitz19}.

\section{Contribution of Rapidly Accreting White Dwarfs}

Rapidly accreting white dwarfs (RAWDs) are white dwarfs in binary systems that accrete mass from a companion star at a rate of about 10$^{-7}$\,$M_\odot$\,yr$^{-1}$ \cite{denissenkov17}. This astrophysical site can synthesize neutron-capture elements via the $i$ process, which is in this case triggered by hydrogen ingestion events \cite{herwig11}. We included the $i$-process yields described in \citeasnoun{denissenkov19} in the one-zone galactic chemical evolution code \texttt{OMEGA} \cite{cote17a} to predict the contribution of RAWDs on the solar abundance distribution. To set the rate of RAWDs in our model, we used predictions generated by the binary population synthesis code \texttt{StarTrack} \cite{Belczynski02,Belczynski08}. We refer to \citeasnoun{cote18_iprocess} for more details on this study.

\begin{figure*}
\center
\includegraphics[width=4.5in]{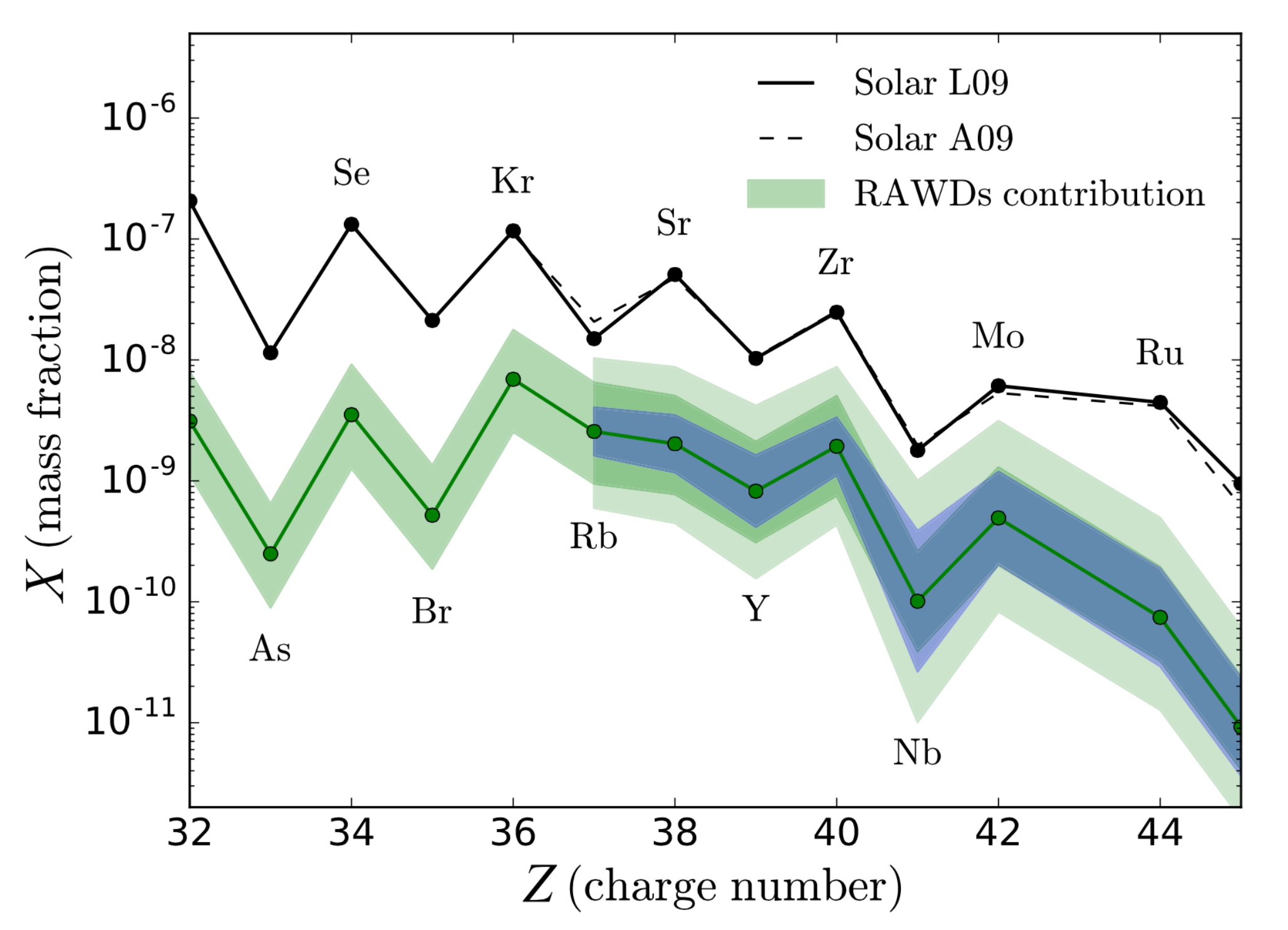}
\captionsetup{justification=justified,singlelinecheck=false,font=small}
\caption{Previously published in \protect\citeasnoun[see Figure~9, \textcopyright AAS, reproduced with permission]{cote18_iprocess}. Predicted contribution of rapidly accreting white dwarfs (RAWDs, green line with markers) against the solar abundance distribution (black lines with markers) of \protect\citeasnoun[A09]{asplund09} and \protect\citeasnoun[L09]{lodders09} for first-peak neutron-capture elements. The dark green and blue bands show the uncertainty generated by galactic evolution and ($n$,$\gamma$) nuclear reaction rate uncertainties, respectively. The light green band shows the combined uncertainty.}
\label{fig_iprocess}
\end{figure*}

As shown in Figure~\ref{fig_iprocess}, RAWDs could have a non-negligible contribution to the origin of first-peak neutron-capture elements in the Sun, in particular for Rb, Sr, Y, Zr, Nb, and Mo. However, the uncertainty in the chemical evolution predictions are currently too large to confidently quantify the role of RAWDs in the production of those elements. The dark green band shows the uncertainty generated by adopting different scenarios for the evolution of metallicity in our Milky Way model. Those scenarios have a significant impact on our predictions because of the strong metallicity-dependence of RAWD yields. The blue band shows the impact of uncertainties in the ($n$,$\gamma$) cross sections present in the nuclear reaction network of the $i$-process nucleosynthesis calculation \cite{denissenkov18}. The largest uncertainties are found for Nb ($\sigma = 0.58$), Mo ($\sigma = 0.38$), and Ru ($\sigma = 0.40$), with their predicted abundances having strong negative and positive correlations with the ($n$,$\gamma$) rates of $^{95}$Y, $^{92}$Sr, and $^{97}$Zr, respectively. The light green band shows the combined uncertainties and demonstrates that our chemical evolution predictions are uncertain by up to two orders of magnitudes.

\section{Contribution of Neutron Star Mergers}

\begin{figure}
\center
\includegraphics[width=4.5in]{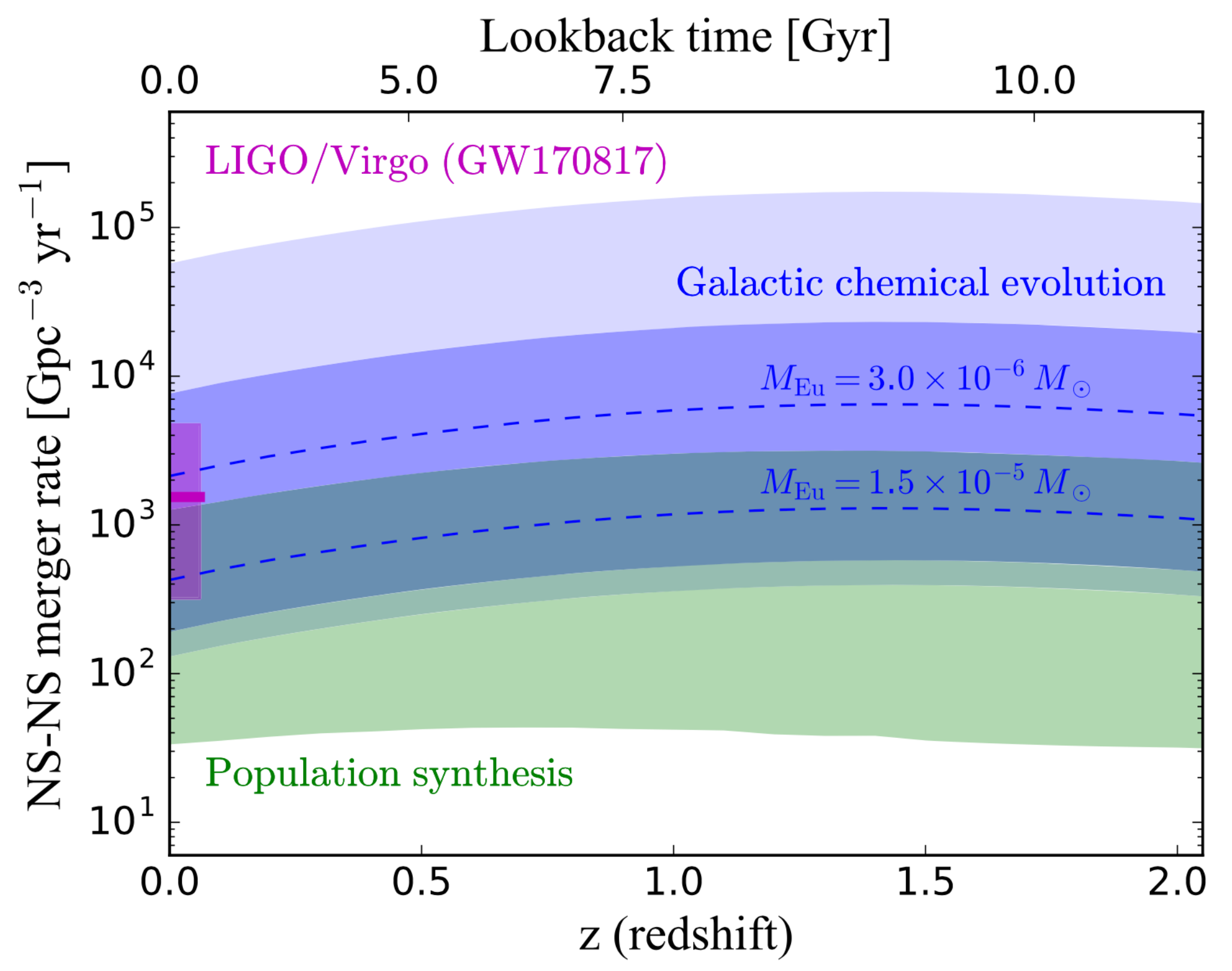}
\captionsetup{justification=justified,singlelinecheck=false,font=small}
\caption{Previously published in \protect\citeasnoun[see Figure~3, \textcopyright AAS, reproduced with permission]{cote18_ligo}. Neutron star merger rate density required by galactic chemical evolution simulations to fit the solar europium (Eu) composition (blue bands), predicted by population synthesis models (green band), and derived from GW170817 by LIGO/Virgo (pink band). The dark blue band shows the uncertainty generated by the variation of results from one chemical evolution code to another, and by the uncertain total mass ejected per neutron star merger, assuming that every merger event ejects the solar $r$-process residuals pattern \protect\cite{arnould07}. The light blue band shows the increased uncertainty when using theoretical $r$-process yields calculations to estimate the mass of Eu ejected per merger event.}
\label{fig_ligo}
\end{figure}

According to numerical simulations, neutron star mergers can synthesize neutron-capture elements via the $r$ process \cite{rosswog99,goriely11,korobkin12,bauswein13,wanajo14,just15,bovard17,lippuner17,radice18,miller19}. The gravitational wave detection GW170817 by LIGO/Virgo \cite{abbott17a} and its electromagnetic emissions confirmed that such events can indeed produce $r$-process elements \cite{abbott17b,cowperthwaite17,kasliwal17,nicholl17,pian17,tanvir17,villar17}. The merger rate density derived by LIGO/Virgo could be high enough for neutron star mergers to be the dominant source of $r$-process elements in the Milky Way \cite{kasen17,cote18_ligo,hotokezaka18,rosswog18}, but many uncertainties remain.

In \citeasnoun{cote18_ligo}, we used the eight chemical evolution simulations compiled in \citeasnoun{cote17b}
to derive the neutron star merger rate required to reproduce the current amount of Eu in the Milky Way, assuming neutron star mergers are the only $r$-process site (see Figure~\ref{fig_ligo}). We chose Eu because it was the element that was common to all eight studies. The dark blue band encompasses two sources of uncertainty: the mass ejected by neutron star mergers, and the variation of predictions from one chemical evolution study to another. For this band, to recover the mass of Eu ejected per neutron star merger, we split the total ejected mass into individual elements assuming the the solar $r$-process residuals elemental distribution \cite{arnould07}. The light blue band shows the increase of uncertainty when using theoretical nucleosynthesis calculations instead of the solar residuals to recover Eu. Those uncertainties were generated by running the $r$-process nucleosynthesis several times using different mass and fission fragment distribution models \cite{mumpower16,vassh19}.

The uncertainty in the merger rate needed by chemical evolution studies is increased by an order of magnitude when including nuclear physics uncertainties. The merger rate derived (or predicted) by any study that uses theoretical $r$-process yields to set the mass of Eu ejected by mergers, is too uncertain to meaningfully quantify the contribution of such events on the $r$-process inventory of the Milky Way. In addition, because theoretical yields do not systematically reproduce the solar $r$-process residuals, using them to match one element (i.e., Eu) with chemical evolution simulations will likely lead to over-productions and under-productions of other $r$-process elements relative to the solar abundances distribution.

We note that binary population synthesis models can independently provide predictions for the neutron star merger rate density \cite[green band in Figure~\ref{fig_ligo}]{chruslinska18}. If neutron star mergers are the dominant source of $r$-process elements in our Galaxy, such predictions should be consistent with the rate required by chemical evolution studies and derived by LIGO/Virgo.

\section{Conclusion}

We reviewed two studies \cite{cote18_iprocess,cote18_ligo} to highlight the impact of nuclear physics uncertainties on the predicted evolution of neutron-capture elements in Milky Way simulations. Nuclear physics uncertainties, including cross-sections in nuclear reaction networks, masses of neutron-rich isotopes, and fission fragment distribution models, induce a significant amount of uncertainty (up to an order of magnitude) in our chemical evolution predictions, which is comparable to the impact galactic evolution uncertainties. In the cases of rapidly accreting white dwarfs and neutron star mergers, our predictions are currently too uncertain to confidently quantify their contribution on the solar elemental composition. Our results serve as a motivation to create and maintain multidisciplinary collaborative efforts to identify and reduce the major sources of uncertainties affecting our interpretation of the origin of the elements.

\newpage 

\noindent{\bf Acknowledgments}

BC acknowledges support from the ERC Consolidator Grant (Hungary) funding scheme (project RADIOSTAR, G.A. n. 724560) and the National Science Foundation (USA) under grant No. PHY-1430152 (JINA Center for the Evolution of the Elements). 
MM was supported by the US Department of Energy through the Los Alamos National Laboratory. Los Alamos National Laboratory is operated by Triad National Security, LLC, for the National Nuclear Security Administration of U.S.\ Department of Energy (Contract No.\ 89233218CNA000001) and by the Laboratory Directed Research and Development program of Los Alamos National Laboratory under project number 20190021DR. 
\\
\\
\noindent{\bf References}
\\
\bibliography{main}

\end{document}